\documentclass[10pt,journal,compsoc]{IEEEtran}

\ifCLASSOPTIONcompsoc
  \usepackage[nocompress]{cite}
\else
  \usepackage{cite}
\fi

\ifCLASSINFOpdf
\else
\fi
\usepackage{amsmath}
\usepackage{hyperref}
\usepackage{graphicx}
\usepackage{multirow}
\usepackage{booktabs}
\usepackage{colortbl}
\usepackage{diagbox}
\usepackage{ragged2e}
\usepackage{algorithm}
\usepackage{algorithmicx}
\usepackage{algpseudocode}
\usepackage[normalem]{ulem}

\usepackage[american]{babel}
\usepackage[numbers,sort&compress]{natbib}

\begin{document}
\definecolor{green}{rgb}{0, 0.5, 0}
\definecolor{orange}{rgb}{0.8, 0.6, 0.2}
\definecolor{red}{rgb}{1.0, 0.0, 0.0}
\definecolor{teal}{rgb}{0.0, 0.4, 0.4}
\definecolor{purple}{rgb}{0.65,0,0.65}
\definecolor{saffron}{rgb}{0.95,0.75,0.2}
\definecolor{turquoise}{rgb}{0.0,0.5,0.5}
\definecolor{black}{rgb}{0.0, 0.0, 0.0}
\definecolor{gray}{rgb}{0.5, 0.5, 0.5}

\newcommand{\chen}[1]{{\color{purple}[Kejiang: #1]}}
\newcommand{\kj}[1]{{\color{blue}#1}}
\newcommand{\hz}[1]{{\color{saffron}[Hang: #1]}}
\newcommand{\hang}[1]{{\color{red}#1}}

\title{Cover Reproducible Steganography via Deep Generative Models}

\author{Kejiang~Chen, 
        Hang~Zhou,
        Yaofei~Wang,
        Menghan~Li,
        Weiming~Zhang,
        Nenghai~Yu

        \IEEEcompsocitemizethanks{
          \IEEEcompsocthanksitem Kejiang~Chen,
        Menghan~Li,
        Weiming~Zhang, and
        Nenghai~Yu are with CAS Key Laboratory of Electro-Magnetic Space Information, School of Cyber Science and Technology, University of Science and Technology of China, Anhui Province Key Laboratory of Cyberspace Security Situation Awareness and Evaluation,
          Hefei 230026, China.(e-mail: chenkj@ustc.edu.cn; zhangwm@ustc.edu.cn.)
            \IEEEcompsocthanksitem     Yaofei~Wang is with School of Computer Science, Hefei University of Technology, China.         
           \IEEEcompsocthanksitem  Hang Zhou is with School of Computer Science, Simon Fraser University, Burnaby, Canada.
          \IEEEcompsocthanksitem Corresponding author: Weiming~Zhang. 
          \IEEEcompsocthanksitem This work was supported in part by the Natural Science Foundation of China under Grant 62102386, 62002334 and 62072421, and by China Postdoctoral Science Foundation under Grant 2021M693091, and by Open Fund of Anhui Province Key Laboratory of Cyberspace Security Situation Awareness and Evaluation.
          }
          \thanks{}

}

\markboth{IEEE Transactions on Dependable Secure Computing}%
{Shell \MakeLowercase{\textit{et al.}}: Bare Demo of IEEEtran.cls for Computer Society Journals}

\IEEEtitleabstractindextext{%
\begin{abstract}
   \justifying
  Whereas cryptography easily arouses attacks by means of encrypting a secret message into a suspicious form, steganography is advantageous for its resilience to attacks by concealing the message in an innocent-looking cover signal. Minimal distortion steganography, one of the mainstream steganography frameworks, embeds messages while minimizing the distortion caused by the modification on the cover elements. Due to the unavailability of the original cover signal for the receiver, message embedding is realized by finding the coset leader of the syndrome function of steganographic codes migrated from channel coding, which is complex and has limited performance. 
  Fortunately, deep generative models and the robust semantic of generated data make it possible for the receiver to perfectly reproduce the cover signal from the stego signal.
  With this advantage, we propose cover-reproducible steganography where the source coding, e.g., arithmetic coding, serves as the steganographic code. Specifically, the decoding process of arithmetic coding is used for message embedding and its encoding process is regarded as message extraction. Taking text-to-speech and text-to-image synthesis tasks as two examples, we illustrate the feasibility of cover-reproducible steganography.
  Steganalysis experiments and theoretical analysis are conducted to demonstrate that the proposed methods outperform the existing methods in most cases. 
\end{abstract}

}

\maketitle

\IEEEdisplaynontitleabstractindextext

\IEEEpeerreviewmaketitle

\IEEEraisesectionheading{\section{Introduction}\label{sec:introduction}}

\IEEEPARstart{S}{teganography} is the art of covert communication that hides secret messages in innocent-looking media. The security of steganography is implicitly built on behavior security that a digital medium that is popular on the Internet could  appropriately serve as the cover medium for message embedding. 
Texts, images, audios and videos have been widely utilized in our daily lives, and their corresponding steganographic algorithms have been well developed in recent years. Among them, steganography algorithms can be divided into three categories: cover selection, cover modification, and cover synthesis steganography~\cite{fridrich2009steganography}. Currently, most attention is focused on cover modification based steganography, which embeds messages by modifying the elements of cover objects (spatial pixels~\cite{HILL,MiPOD}, chaotic pixels~\cite{valandar2017new,valandar2019integer}, audio waveform~\cite{aacstc,chen2019defining}, etc.), yet such modification will ineluctably induce distortion, which is likely to be exposed to steganalysis, the opposite of steganography. 

Following the timeline of steganography, cover-modification steganography has developed from constant distortion, wet distortion  to arbitrary distortion. For constant-distortion steganography, the costs of modifying different elements in cover objects are the same, which is equivalent to minimizing the modification number of elements. 
Many steganographic codes for constant-distortion steganography 
~\cite{westfeld2000high,van2001embedding,schonfeld2006embedding,zhang2009fast} achieve considerable performance. However, it is found that some elements of cover objects cannot be modified, such as the saturated area in spatial images. To solve the above problem, wet codes are proposed, and some methods~\cite{fridrich2005efficient,zhang2008maximizing} have approached the theoretical bound of the rate-distortion function. When diving deeper into the research, it can be inferred that it is more reasonable to assign small modification costs to complex areas of the cover object, which corresponds to adaptive steganography, the mainstream research. To minimize the arbitrary additive distortion of the modification, two representative steganographic codes, Syndrome Trellis Codes (STCs)~\cite{STC} and Steganographic Polar Codes (SPCs)~\cite{li2020designing} are proposed with the performance asymptotically achieving theoretical bounds. With the well-developed steganographic codes, many classical distortion-based methods have emerged, such as~\cite{WOW,UNIWARD,HILL,MS} in image,~\cite{aacstc,chen2019defining,wu2020audio} in audio, and~\cite{yao2015defining,zhai2019universal,chen2021ddca} in video. However, due to the utilization of the near-optimal steganographic codes, regardless  of how well the distortion is defined, there is still room for improving the security performance. 

Recently, deep generative models have been considerately developed, which provide a way to sample new data from a distribution learned from training data. Gartner predicts that by 2025 generative AI will account for 10\% of all data produced~\cite{Gartner2022}. In addition, the metaverse has become a hotspot, which is also full of generative data~\cite{lee2021all}, indicating that generative data are suitable for steganography. More importantly, generative models largely change the data environment of steganography. Specifically, by utilizing the same generative models, the receiver can reproduce the same cover signal with the same semantic information, which motivates us to design new steganographic methods.

In this paper, we first briefly review the motivation of designing steganographic codes for encoding messages into cover objects, which 
are originally derived from channel coding, and summarize their limitations. 
The underlying reason is that the receiver has no access to the original cover, so the syndrome function is used for message extraction, where the coset leader of the syndrome function is adopted to guide the modification.
Inspired by the signal reproducibility of deep generative models, we propose cover-reproducible steganography, a source coding based coding scheme (e.g., arithmetic coding) that performs message embedding and extraction, and validate the approach on two popular generative tasks: text-to-speech and text-to-image. 
The experimental results demonstrate that the proposed methods outperform the STC
 and SPC-based steganography methods in most cases.


In the remainder of this paper, Section~\ref{sec2} introduces the related work of minimizing distortion steganography. In Section~\ref{sec3}, we prove that the distortion sort scheme does not achieve better security performance. In Section~\ref{sec4}, 
we present the framework of cover reproducible steganography and two instances based on text-to-speech and text-to-image synthesis tasks. Experimental results and analysis are elaborated in Section~\ref{sec5}. The following part, Section~\ref{sec6}, concludes the paper.

\section{Preliminaries and Related Work}
\label{sec2}
In this paper, matrices, vectors and sets are written in bold-face. The cover sequence is denoted by $\mathbf{x} = (x_1, x_2, . . . , x_n)$, where the signal $x_i$ is an integer, such as the value of a sample in an audio clip. The modification pattern on $x_i$ is formulated by the range $I$. For example, the $\pm 1$ embedding operation is ternary embedding with $I=\{-1,0,+1\}$, where $0$ denotes no modification.

\subsection{Minimizing Distortion Steganography}
In  adaptive steganography, the elements in different regions will be assigned different costs.  Given a cover object $\mathbf{x}$, the cost introduced by modifying $x_i$ to $y_i$ can be denoted by $\rho_i$. In regard to additive steganography, the distortion is the sum of all costs, $D(\mathbf{x},\mathbf{y})=\sum_{i=1}^{n} \rho_i,\ \mathbf{y}\in \mathcal{Y} $. The modification probability is denoted as $\pi({\mathbf{x},\mathbf{y}})$, thus the sender can send up $H\left(\pi({\mathbf{x},\mathbf{y}})\right)$ bits of message with the expected value of the distortion $E_\pi(D)$, where
\begin{equation}
     H\left(\pi({\mathbf{x},\mathbf{y}})\right)=-\pi\left(\mathbf{x},\mathbf{y}\right)\log_2\pi\left(\mathbf{x},\mathbf{y}\right),
\end{equation}
\begin{equation}
    E_\pi(D)= \sum_{\mathbf{y} \in \mathcal{Y}} \pi(\mathbf{x},\mathbf{y})D(\mathbf{x},\mathbf{y}).
\end{equation}
For a given message length $L$, minimizing the distortion while embedding message can be formulated as the following optimization problem:
\begin{equation}
\begin{aligned}
\min_{\pi} \quad & E_\pi(D)\\
\textrm{s.t.} \quad & H\left(\pi({\mathbf{x},\mathbf{y}})\right)=L.
\end{aligned}
\end{equation}
This problem can be solved using Lagrange multipliers. The optimal probability $\boldsymbol{\pi}_\lambda$ follows an exponential distribution with respect to $D(\mathbf{x},\mathbf{y})$:
\begin{equation}
     \boldsymbol{\pi}_\lambda = \frac{1}{Z(\lambda)}\exp\left({-\lambda D(\mathbf{x},\mathbf{y})}\right),
\label{gibbs}
\end{equation}
where $Z(\lambda)$ is a normalizing factor:
\begin{equation}
    Z(\lambda) = \sum_{y \in \mathcal{Y}}\exp \left({-\lambda D(\mathbf{x},\mathbf{y})}\right),
\end{equation}
and $\lambda$ is the Lagrange multiplier determined from the message length constraint. As proven in~\cite{Gibbs}, the entropy is decreasing in $\lambda$, so $\lambda$ can be quickly determined by binary search. For additive steganography, the optimal $\boldsymbol{\pi}_\lambda$ is given by 
\begin{equation}
    \pi_{\lambda}(x_i,y_i)=\frac{\exp(-\lambda\rho_i(x_i,y_i))}{\sum_{y_i \in I+x_i} \exp(-\lambda\rho_i(x_i,y_i))}.
    \label{eq-optimal}
\end{equation}
The optimality of $\pi_{\lambda}$ implies that $E_\pi(D)$ of any probability distribution $\pi$ satisfying (3) cannot be smaller than $E_{\pi_{\lambda(D)}}$, i.e.,
\begin{equation}
    E_{\boldsymbol{\pi}}(D) \geq E_{\boldsymbol{\pi}_{\lambda}}(D),
\end{equation}
with equality iff $\boldsymbol{\pi}=\boldsymbol{\pi}_{\lambda}$.

\subsection{Steganographic Codes}
For additive steganography, there exist practical coding methods, such as STCs~\cite{STC} and SPCs~\cite{li2020designing}, which can approach the lower bound of the average distortion. These codes are derived from the channel coding, to solve the problem that the recipient cannot obtain the cover. The syndrome function Eq.  (\ref{syndromefunction}) is used for message extraction, and finding the coset leader of the function is equivalent to embedding message with near-minimum distortion:
\begin{equation}
Ext(\mathbf{y})= \mathbf{H}\mathbf{y}=\mathbf{m}\xrightarrow{\mathbf{y}=\mathbf{x}+\mathbf{e}} \mathbf{H(x+e)}=\mathbf{m},
\label{syndromefunction}
\end{equation}
\begin{equation}
    Emb(\mathbf{x},\mathbf{m})=\arg \min_{\mathbf{y} \in C(\mathbf{m})} D(\mathbf{x},\mathbf{y}),
\end{equation}
where $\mathbf{e}$ is the modification pattern corresponding to near-minimum distortion.

To find $\mathbf{e}$ more effectively, the parity check matrix $\mathbf{H}$ is carefully designed in STCs, which is constructed by placing a small $h \times w$ submatrix $\hat{\mathbf{H}}$ along the main diagonal. The solution of STCs can be represented as a path through the syndrome trellis of  $\mathbf{H}$. The height $h$ of the submatrix determines the number of paths, which affects the algorithm speed and efficiency.
There are $kh$ choices in each grid of the trellis for $k$-ary embedding. Therefore, a larger $h$ means a more powerful ability to minimize distortion but also higher computational complexity. Besides, the complexity of STCs  exponentially increases with the number of modification patterns $k = |I|$.

SPCs provide another near-optimal steganographic coding method based on polar codes, using Successive Cancellation List (SCL) decoding algorithm to minimize additive distortion in steganography. Arikan's efficient method~\cite{arikan2009channel} is used to determine the frozen indices $ \mathcal{A}^{c} $ for constructing the steganographic parity-check matrix $\mathbf{H}$. The list size $l$ of SCL determines the performance of the coding, and a larger $l$ leads to a better ability to minimize distortion but also higher computational complexity.

\subsection{Generative Models}
Deep generative models are neural networks with many hidden layers trained to approximate complicated, high-dimensional probability distributions using a large number of samples~\cite{ruthotto2021introduction}. Text-to-speech and text-to-image are two representative tasks.

\subsubsection{Text-to-Speech (TTS)}
Deep generative models have proven to be effective in producing natural speech. Modern TTS systems often consist of two parts: the first part converts the input text into acoustic features (feature generator), and the second one synthesizes the raw waveform conditioned on these features (vocoder). Among feature generators, Tacotron2~\cite{shen2017natural}, Transformer-TTS~\cite{li2019neural} and Flowtron~\cite{flowtron} enabled highly natural speech synthesis. Producing acoustic features frame by frame, they achieve considerable mel-spectrogram reconstruction from the input text. WaveNet~\cite{oord2016wavenet}, WaveGlow~\cite{prenger2019waveglow}, MelGAN~\cite{kumar2019melgan} and HiFi-GAN~\cite{kong2020hifi} are widely used as vocoders, which generate raw waveforms with high voice quality. It is noteworthy that some TTS systems output the same speech with the same input text, which provides us with an opportunity to make it possible that the receiver to obtain the cover speech. 

\subsubsection{Text-to-Image (TTI)}
\label{sec-TTS}
Deep neural networks based on Generative Adversarial Networks (GANs)~\cite{reed2016generative} have enabled end-to-end trainable text-to-image generation. To enable TTI models to synthesize higher resolution images, many following works are proposed to use multiple, stacked generators, such as StackGAN~\cite{zhang2017stackgan} and its enhanced version StackGAN++~\cite{zhang2018stackgan++}. Attention has a major impact on improving language and vision tasks, and AttnGAN~\cite{xu2018attngan} builds upon StackGAN++ and incorporates attention into a multi-stage refinement pipeline. The transformer-based TTI model, DALL-E~\cite{ramesh2021zero} has also been proposed with considerable performance. Inspired by CycleGAN, cycle-consistent image generation by appending an image caption network and training the generation network to produce a synthesized image with similar caption as the text, such as MirrorGAN~\cite{qiao2019mirrorgan} and N2N~\cite{rombach2020network}. The reversibility of cycle-consistent TTI makes the information hiding and extraction easier to implement.

\section{Lossy Property of Distortion-Sort Steganography}
\label{sec3}
Generative models, such as text-to-speech and text-to-image systems, enable the situation in which the receiver owns the cover object. 
With the cover audio, the first thought that comes to mind is that the sender can sort the cover elements according to the distortion and embed message into a part of cover elements with minimal distortion. In the original setting, for the message of $m$ bits and the cover sequence $\mathbf{x} = (x_1, x_2, \cdots , x_n)$ with cost $\rho({\mathbf{x}}) = (\rho_{x_1},\rho_{x_2},\cdots,\rho_{x_n})$, 
the average distortion introduced by message embedding can be minimized when the probability distribution $\mathbf{\pi}$ follows
Gibbs distribution (\ref{gibbs}). In this manner, the minimal average
distortion is computed by 
\begin{equation}
 E_{\boldsymbol{\pi}_{\lambda}}\left(D_{\mathbf{x}}\right)=\sum_{i=1}^{n} \pi_{\lambda}\left(x_{i}\right) \rho\left(x_{i}\right), 
\end{equation}
where $\pi_{\lambda}\left(x_{i}\right)=\frac{\exp \left(-\lambda \rho\left(x_{i}\right)\right)}{1+\exp \left(-\lambda \rho\left(x_{i}\right)\right)} $ and $ m=\sum_{i=1}^{n} H\left(\pi_{\lambda}\left(x_{i}\right)\right)$.

In the distortion-sort setting, the cover elements are arranged in ascending order of cost, $\mathbf{x}'=(x'_0,x'_1,\cdots,x'_n), \rho(x'_i) \leq \rho(x'_{i+1})$. In order to embed $m$ bits, the sender marks elements $1,\cdots,k$ as changeable and embeds the message into changeable elements. Similarly, the minimal average distortion of the optimal distribution $\boldsymbol{\mu}_{\lambda^{\prime}}$ with respect to $\rho(\mathbf{x}')$ becomes 
\begin{equation}
    E_{\boldsymbol{\mu}_{\lambda^{\prime}}}\left(D_{\mathbf{s}}\right)=\sum_{i=1}^{k} \mu_{\lambda^{\prime}}\left(x'_{i}\right) \rho\left(x'_{i}\right).
\end{equation}
Since the probability distribution about the complementary set can be regarded as $\mathbf{0}$, we can consider that the distribution $\boldsymbol{\mu}_{\lambda^{\prime}}$ complemented with $\mathbf{0}$ is a distribution about $\rho_{\mathbf{x}'}$, i.e., $ \boldsymbol{\pi}=\left(\boldsymbol{\mu}_{\lambda^{\prime}}. \mathbf{0}\right) $.
From Eq. (\ref{gibbs}), we thus have
\begin{equation}
 E_{\boldsymbol{\mu}_{\lambda^{\prime}}}\left(D_{\mathbf{x}'}\right)>E_{\boldsymbol{\pi}_{\lambda}}\left(D_{\mathbf{x}}\right).
\end{equation}
Therefore, the distortion sort steganography cannot achieve better security performance than the original minimal distortion steganography.

\section{Cover Reproducible Steganography}
\label{sec4}
The generative models provide us with a great opportunity for the receiver to reproduce the cover with the same semantic. In addition, the modification seldom changes the semantic of the generated media, since the modification is very slight with respect to the original signal. Under this circumstance, the modification pattern and the optimal probability can be obtained by the receiver as well. The benefit is that channel coding is no longer needed for message embedding and extraction. The receiver can synchronize the modification pattern with the sender. Therefore, the source coding can be adopted for message embedding, which not only possesses high computational efficiency but also near-optimal rate-distortion performance. Using this property, we propose a novel steganography method cooperating with Adaptive Arithmetic Coding, named Cover Reproducible Steganography (abbreviated as CRS). In Section \ref{message_emb}, we develop the algorithm of message embedding and extraction. In Section \ref{audio_stegano} and Section \ref{image_stegano}, two instances, audio steganography and image steganography, are presented. 

\subsection{Message Embedding and Extraction}\label{message_emb}

\textbf{1) Message embedding:}
Given the cover $\mathbf{x}$ and the corresponding distortion definition method $D()$, we can obtain the cost $\rho$ of changing every element. Following the optimal distribution, given the message length $L$ and the cost $\boldsymbol{\rho}$, the optimal probability $\boldsymbol{\pi}$ is obtained according to Eq. (\ref{eq-optimal}). With the probability $\boldsymbol{\pi}$, we can decode the encrypted message into modification pattern $\mathbf{e}$ using the decoding process of arithmetic coding.

\begin{figure}
	\centering
	\includegraphics[width=3.45in]{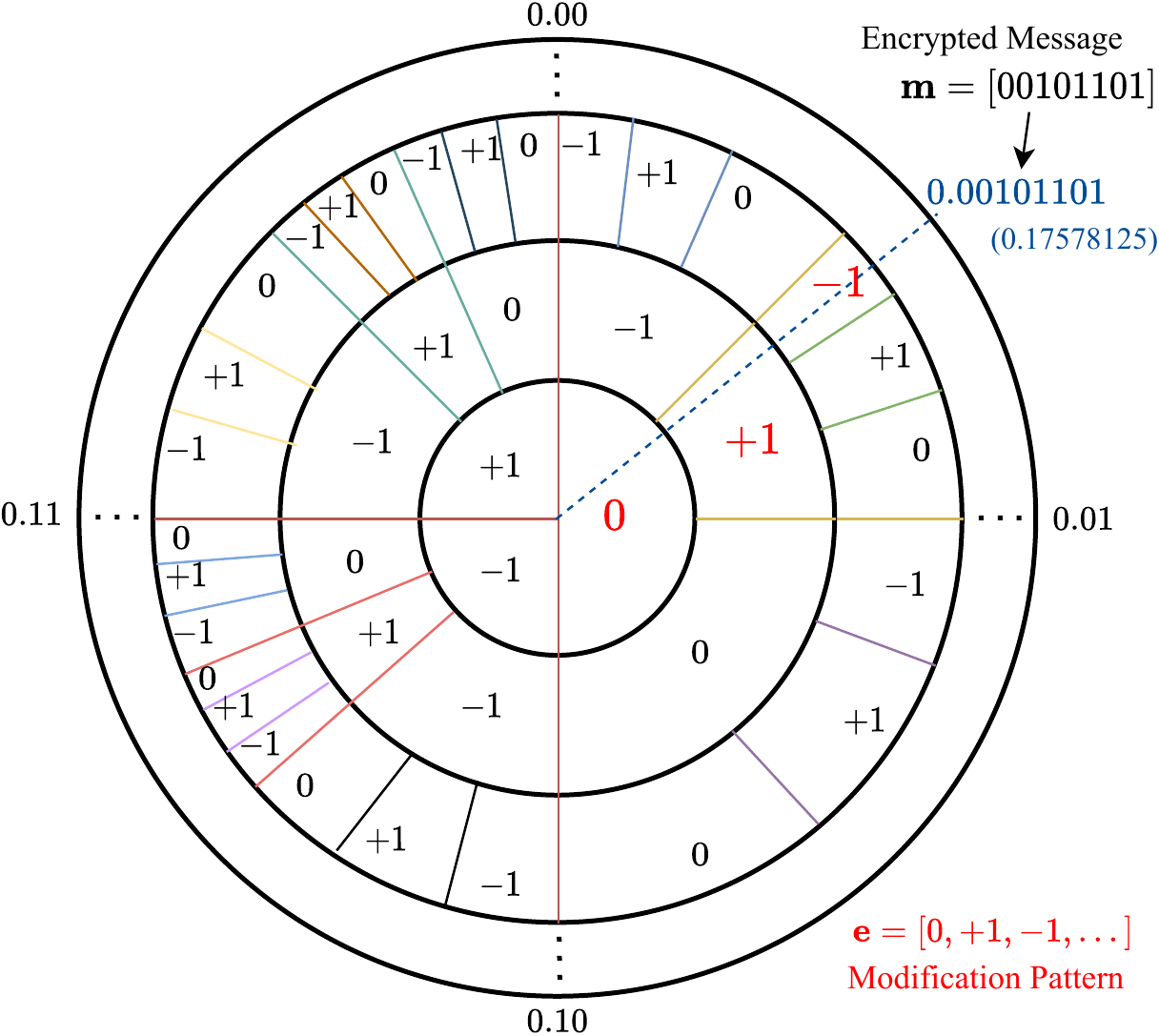}
	\caption{Example of arithmetic coding for ternary steganography. Concentric circles represent sample index; the innermost represents $i = 1$, the middle $i=2$, and the outer $i = 3$. Each circle represents the conditional distribution $\pi(x_i)$. 
		The encrypted secret message $\mathbf{m}=[00101101]$ is viewed as a binary representation of a fraction $0.00101101$ in the range $[0, 1)$, then the modification pattern $\mathbf{e}$ is determined on the route where the fraction lies in.
		Message extraction is performed via the reverse operation: the sequence of modification patterns progressively narrows the range of possible messages.
	}
	\label{fig:arithmetic}
\end{figure}

Arithmetic coding maps a string of elements to a uniformly distributed binary string. Here, the decoding process of arithmetic coding can be used for message embedding: first the uniform message is selected, and then the message is mapped to a sequence (modification pattern). Following the previous work~\cite{ziegler2019neural}, the message embedding can be represented by finding a path in the concentric circle, as shown in Figure~\ref{fig:arithmetic}. Concentric circles represent samples (pixels in images, waveform samples in audio clips). The innermost circle represents $x_1$, the middle $x_2$, and so on. The intervals in each circle represent the probabilities of modification patterns.

To encode the secret message into a modification pattern, the secret message $\mathbf{m}$ is viewed as a binary representation of a fraction in the range $[0, 1)$. The secret message $\mathbf{m}$ is first encrypted using the XOR operation with a pseudo random bitstream $\mathbf{s}$
\begin{equation}
    \mathbf{m}_e = \mathbf{m} \oplus \mathbf{s},
\end{equation}
where $\oplus$ represents the XOR operation. Given the encrypted message $\mathbf{m}_e=[m_1m_2m_3...m_L]$, it can be interpreted as a fraction $q$ in the range $[0,1)$ by prepending ``0.'' to it:
\begin{equation}
m_1m_2m_3...m_L  \rightarrow q = 0.m_1m_2m_3...m_L=\sum_{i=1}^L m_i \cdot 2^{-i}.
\label{fraction}
\end{equation}
The fraction $q$ uniquely marks a point on the edge of the circle, as well as a line from the origin to the point. Message embedding is carried out by simply reading off the modification patterns corresponding to the bins. The embedding process stops when the list of modification patterns unambiguously defines the message. The stego $\mathbf{y}$ is easily obtained by adding the modification $\mathbf{e}$ to the cover $\mathbf{x}$.
\begin{equation}
    \mathbf{y} = \mathbf{x} + \mathbf{e}.
\end{equation}
The stego $\mathbf{y}$ are sent to the receiver.
The message length $L$ can be negotiated in advance.

\textbf{2) Message extraction:}
Receiving the text and the corresponding stego $\mathbf{y}$, the receiver first reproduces the cover $\mathbf{x}$, then he can get the probability distribution $\boldsymbol{\pi}$ and the modification pattern $\mathbf{e}$. With the same probability distribution $\boldsymbol{\pi}$, the same concentric circles can be reconstructed. Message extraction is performed via the reverse operation: the modification pattern progressively narrows the range of possible messages until there only exists one fraction $q = \sum_{i=1}^L m_{i}2^{-i}$ of $L$-length message $\mathbf{m}_e$ in the interval. 

With the shared secret key, the receiver can generate the same pseudo-random binary string $\mathbf{s}$ and then can decrypt the message:
\begin{equation}
    \mathbf{m} = \mathbf{m}_e \oplus \mathbf{s}.
\end{equation}

Figure~\ref{fig:arithmetic} gives an example of message embedding and extraction. Concentric circles represent the sample index; the innermost represents $i = 1$, the middle $i = 2$, and the outer $i = 3$. Each circle represents the conditional distribution $\pi(x_i)$, e.g., $\pi^{0}(x_1)=0.5, \pi^{-1}(x_1)=0.25$ and $\pi^{+1}(x_1)=0.25$. The encrypted secret message $\mathbf{m}=[00101101]$ is viewed as a binary representation of a fraction $0.00101101$ $(0.17578125)$ in the range $[0, 1)$, then the modification pattern $\mathbf{e}$ is determined by the route where the fraction lies in. Message extraction is performed via the reverse operation: the sequence of modification patterns progressively narrows the range of possible messages.

The previous processes only consider a short message. When the message becomes longer, due to the precision limitation (32 bits or 64 bits), a queue-based message embedding mechanism is introduced. The details are illustrated in 
Algorithm \ref{aad} and Algorithm \ref{algaac}. 
\begin{algorithm}[H]
	{}\caption{Message embedding}
	\label{aad}
	\begin{algorithmic}[1]
		\Require The cover $\mathbf{x}$, the encrypted message $\mathbf{m}_e$, the optimal probability distribution $\boldsymbol{\pi}$, the precision $\beta$.
		\Ensure The stego $\mathbf{y}$.
		\State $h_0 = 1,l_0 = 0,p=0$
		\For{$k \in \{ 1,2,3,...,n\}$}
		\For {$j \in \{ -1,0,+1\}$}
		\State $q = 0.m_pm_{p+1}m_{p+2}...m_{p+\beta}$   \Comment{Queue-based Message}
		\State $h_k = l_{k-1}+(h_{k-1}-l_{k-1})*\sum_{i=-2}^{j} \pi^i\left(x_k\right)$\Comment{$\pi^{-2}=0$} 
		\State $l_k = l_{k-1}+(h_{k-1}-l_{k-1})*\sum_{i=-2}^{j-1} \pi^i\left(x_k\right)$  
		\If {$q \in [l_k,h_k)$}
		\State $\mathbf{e} = \mathbf{e} :: j$
		\State $\mathcal{R} = \{ r=0.b_1b_2b_3...b_\beta, b\in \{0,1\}| r\in [l_k,h_k) \}$  
		\State \textbf{break}
		\EndIf
		\EndFor 
		\For{$i \in \{1,...,\beta \}$} \Comment{Calculate the actual embedded message length.}
		\If {$r_1[b_i]=r_2[b_i]=\cdots=r_{|\mathcal{R}|}[b_i]$}
		\Else
		\State $L' =i-1$ 
		\State \textbf{break} 
		\EndIf
		\EndFor
		\State $l_k = l_k \ll L', h_k =  (h_k-2^{-\beta} ) \ll L'+ 2^{-L'} $  \Comment{ $\ll$ Bit left shift operation of fraction part.}
		\State $p = p+L'$
		
		\EndFor
		
		\State $\mathbf{y=x+e}$
		\State $output = \mathbf{y}$	\end{algorithmic}
\end{algorithm}

\begin{algorithm}[H]
	{}\caption{Message extraction}
	\label{algaac}
	\begin{algorithmic}[1]
		\Require The stego $\mathbf{y}$, the reproduced cover  $\mathbf{x}$, the optimal probability distribution $\boldsymbol{\pi}$, the precision $\beta$.
		\Ensure The  encrypted message $\mathbf{m}_e$.
		\State $\mathbf{e=y-x}$
		\State $h_0 = 1,l_0 = 0,p=0,\mathbf{m}_e=\{\}$
		\For{$k \in \{ 1,2,3,...,n\}$}
		\State $h_k = l_{k-1}+(h_{k-1}-l_{k-1})*\sum_{i=-2}^{e_k} \pi^i\left(x_k\right)$\Comment{$\pi^{-2}=0$} 
		\State $l_k = l_{k-1}+(h_{k-1}-l_{k-1})*\sum_{i=-2}^{e_k-1} \pi^i\left(x_k\right)$
		\State $\mathcal{R} = \{ r=0.b_1b_2b_3...b_\beta, b\in \{0,1\}| r\in [l_k,h_k) \}$ 
		\For{$i \in \{1,...,\beta \}$}    \Comment{Calculate the actual embedded message length.}
		\If {$r_1[b_i]=r_2[b_i]=\cdots=r_{|\mathcal{R}|}[b_i]$}
		\Else
		\State $L' =i-1$ 
		\State \textbf{break} 
		\EndIf
		\EndFor
		\State $l_k = l_k \ll L', h_k = ( h_k-2^{-\beta} ) \ll L'+ 2^{-L'} $ \Comment{ $\ll$  Bit left shift operation of fraction part.}
		\State $\mathbf{m}_e = \mathbf{m}_e :: r_1[1:L']$
		\EndFor
		\State $output = \mathbf{m}_e$	\end{algorithmic}
\end{algorithm}

\begin{figure*}
    \centering
    \includegraphics[width=6.5in]{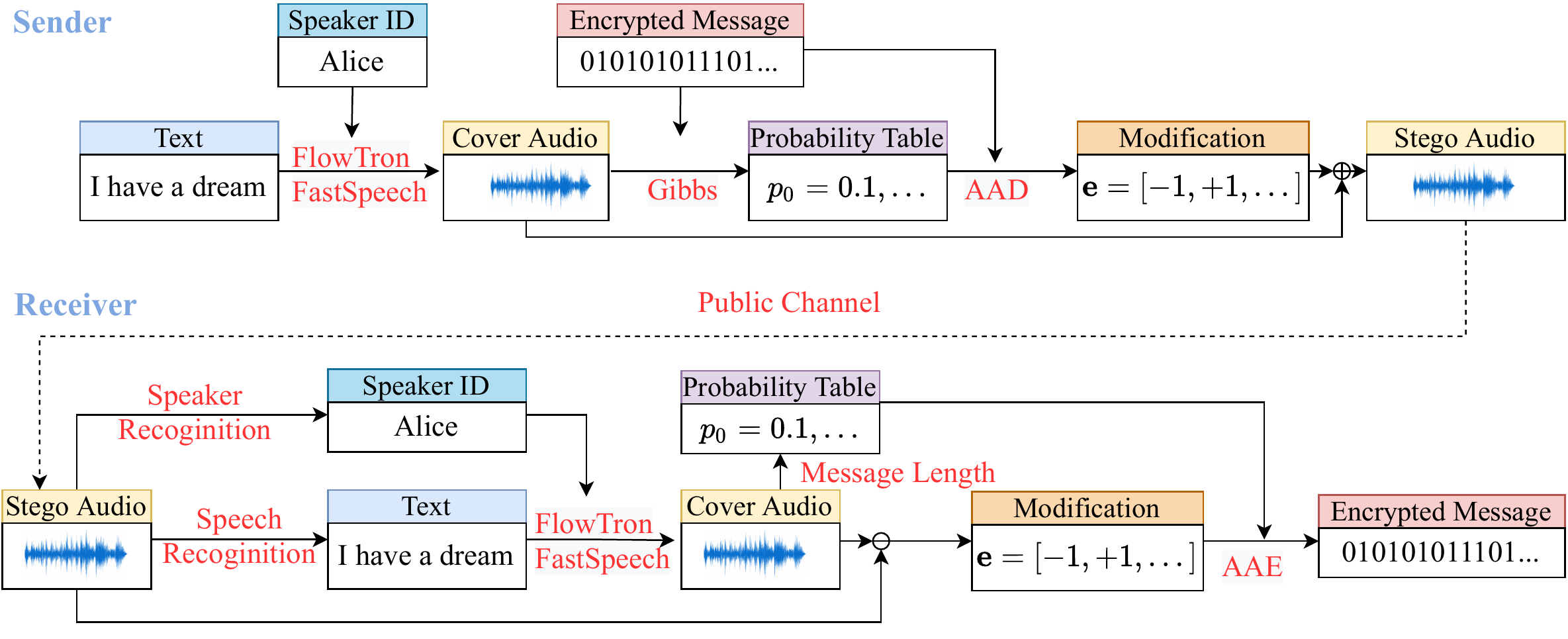}
    \caption{Cover Reproducible Steganography based on text-to-speech system.
    }
    \label{fig:crs_tts}
\end{figure*}

\subsection{The Proof of Near-optimal Performance}\label{theoreticalbound}
Information embedding and source coding (compression) are dual problems~\cite{DBLP:journals/tit/BarronCW03}. Intuitively, the embedding performance of using a certain compressor is equivalent to its compression performance due to their duality~\cite{DBLP:journals/tit/BarronCW03}. 
In our scenario, if the compression is perfect, then the modification pattern will strictly follow the optimal modification probability distribution. However, there is no perfect compressor, and we choose near-optimal Arithmetic Coding for message embedding. Inevitably, the probability  distribution that actually follows differs from the optimal probability distribution. According to the compression performance of Arithmetic coding, we can derive the upper bound of the difference of the real distribution and the optimal distribution. 

Given the optimal modification probability distribution $ \boldsymbol{\pi} $ of length $n$, the bounds on the length of an arithmetic code to represent the distribution are~\cite[Section 4.4.1]{sayood2017introduction}:
\begin{equation}
	H(\mathbf{\boldsymbol{\pi}}) \le L < H(\mathbf{\boldsymbol{\pi}})+\frac{2}{n}.
	\label{arith_bound}
\end{equation}

In practice, we embed only $ H\left(\boldsymbol{\pi}\right) $ bits of message, so the real distribution  arithmetic code used for message embedding must be modified to $\boldsymbol{\pi}'$  except for the situation $ L=H\left(\boldsymbol{\pi}\right)$. According to~\cite[Theorem 5.4.3]{cover2012elements}, using the incorrect distribution $\boldsymbol{\pi}'$ for encoding  $H(\boldsymbol{\pi})$ bits message  when the target distribution is $\boldsymbol{\pi}$ incurs a penalty of $D\left( \boldsymbol{\pi} \parallel \boldsymbol{\pi}'\right)$ in the average description length. 
Directly extended from Eq. (\ref{arith_bound}), the distribution divergence $D\left( \boldsymbol{\pi} \parallel \boldsymbol{\pi}'\right)$ has an upper bound:
\begin{equation}
	D\left( \boldsymbol{\pi} \parallel \boldsymbol{\pi}'\right)<\frac{2}{n},
	\label{dupper}
\end{equation}
and if $n \rightarrow \infty$, then:
\begin{equation}
	D\left( \boldsymbol{\pi} \parallel \boldsymbol{\pi}'\right) \rightarrow 0.
	\label{dupper2}
\end{equation}
By increasing the length of the sequence, the relative entropy between $\boldsymbol{\pi}$ and $\boldsymbol{\pi}'$ becomes 0, meaning that the modification using arithmetic coding follows the optimal modification probability distribution $\boldsymbol{\pi}$ when the sequence is long enough.

\subsection{Text-to-Speech CRS}\label{audio_stegano}
Text-to-speech systems have developed rapidly and the generated speeches have excellent quality. More importantly, these systems facilitate efficiently performing CRS. Figure~\ref{fig:crs_tts} shows the diagram of cover reproducible steganography based on the TTS system. At the sender-end, the text is fed into the TTS system generating the cover audio. With the cover audio and the secret message, we can decode the encrypted message into the modification pattern using adaptive arithmetic decoding according to the optimal probability distribution. Then the stego audio is obtained by adding the modification pattern to the cover audio. 

At the receiver-end, the stego audio is first recognized as text, which is the same as that at the sender-end. It is worth mentioning that the semantic of the audio is robust to the modification, since the modification is slight with respect to the original audio signal. Notably, someone may argue that the speech recognition system could not achieve the perfect performance. At first, the speech recognition system has achieved considerable performance, such as TDNN~\cite{xiang2019margin} for speaker recognition and DeepSpeech~\cite{DBLP:conf/icml/AmodeiABCCCCCCD16} for text recognition. Furthermore, the process can be checked by the sender, if the recognition failed, the sender can change the text using synonym substitution.
In addition, the receivers themselves can also help to recognize the speaker and the speech content.
Thereafter, using the same TTS models, the cover audio and the modification pattern can be reproduced. According to the negotiated distortion function, the identical optimal probability distribution is obtained by inputting the message length. The encrypted message can be extracted using adaptive arithmetic encoding. For the multi-speaker situation, a speaker recognition model is trained and shared, so that the receiver can identify the speaker and carry out the same generation process.

\begin{figure*}
    \centering
    \includegraphics[width=6.5in]{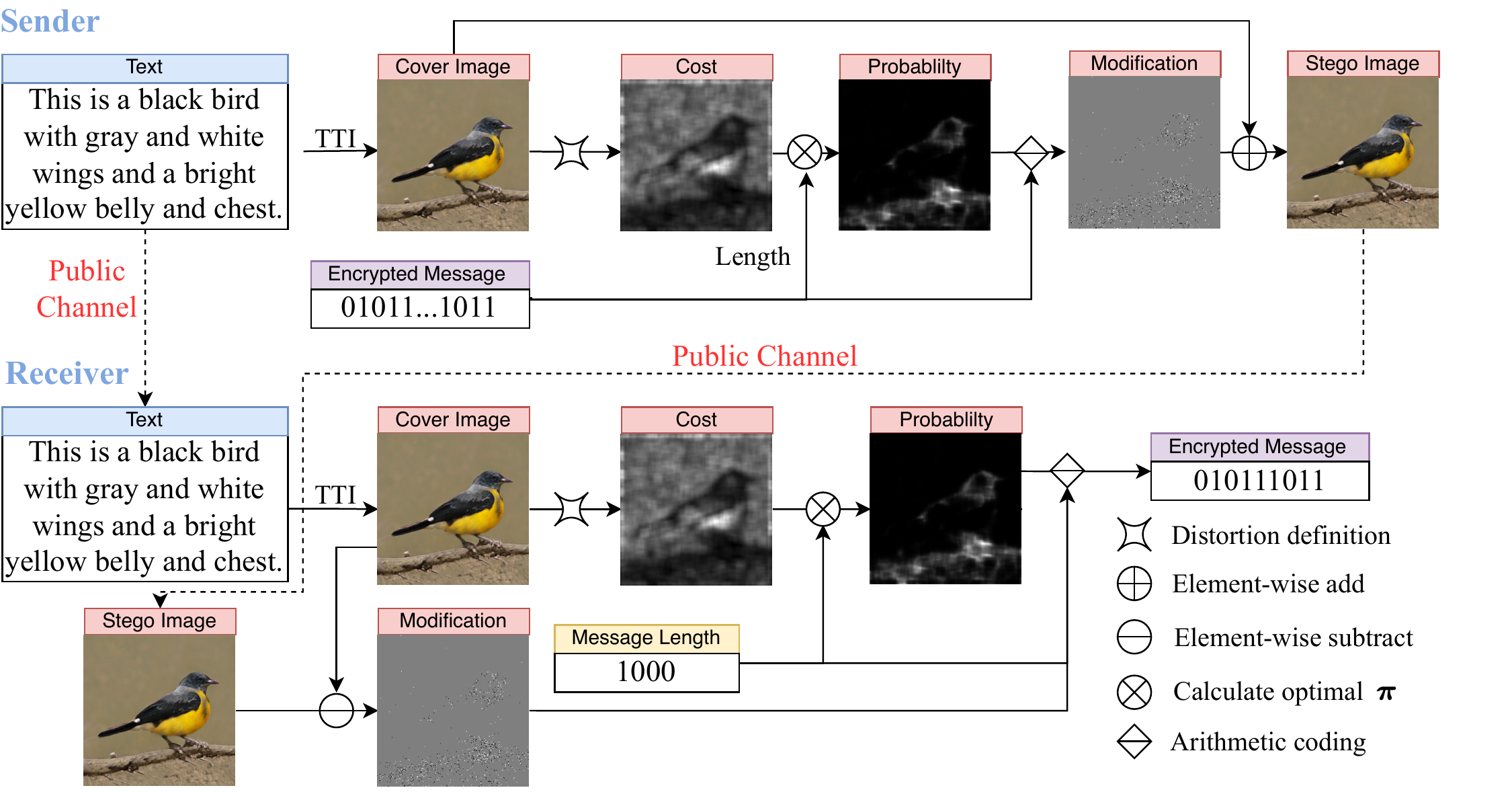}
    \caption{Cover Reproducible Steganography based on text-to-image system.
    }
    \label{fig:crs_tti}
\end{figure*}
\subsection{Text-to-Image CRS}\label{image_stegano}
Sharing images with a caption is very popular on social websites, meaning that it is a suitable way for covert communication. Besides, text-to-image generation has gradually turned into practical and is suitable for cover reproducible steganography. Figure~\ref{fig:crs_tti} shows the diagram of cover reproducible steganography based on the TTI system.
At the send-end, the text is fed into TTI system generating the cover image. The modification costs of image pixels are defined using image steganography distortion function, such as  HILL~\cite{HILL} and MiPOD~\cite{MiPOD}. According to  minimal distortion steganography, by feeding the message length and costs, we can obtain the optimal probability distribution. Subsequently, the encrypted message is decoded into the modification patterns using adaptive arithmetic decoding. The stego image is obtained by adding the modification patterns to the cover image. The stego image and the text are posted on the website together. At the receiver-end, the text is fed into the same TTI system generating the cover image. Naturally, the modification pattern is obtained. Using the negotiated distortion function and message length, the optimal probability distribution is calculated. Thereafter, the encrypted message is extracted by encoding the modification patterns according to the probability distribution. After decryption, the receiver obtains the secret message.

Additionally, as mentioned in Section~\ref{sec-TTS}, the TTI also has cycle-structure generators, e.g., MirrorGAN, which can perform text-to-image as well as image-to-text.
Under this circumstance, the CRS framework can be applied to the scenario where only a single image is sent, because the modification seldom changes the semantic of image and the stego image can be captioned into the same text.

\section{Experiments}
\label{sec5}
In this section, experimental results and analysis are presented to demonstrate the feasibility and effectiveness of the proposed schemes.
\subsection{Experimental Setting}
\subsubsection{Dataset and Generative Models}
\textbf{Text-to-Speech:} For single-speaker, the Flowtron~\cite{flowtron} and WaveGlow~\cite{prenger2019waveglow} are trained on LJ speech~\cite{ljspeech172} dataset, which consists of 13,100 short audio clips of a single speaker reading passages from 7 nonfiction books. The dataset consists of approximately 24 hours of speech data recorded on a MacBookPro using its built-in microphone in a home environment. The sampling rate of audio is set as 22.05 kHz. With the well-trained generative models, given 10,000 random sentences, we generate the corresponding 10,000 audio clips, whose bits per sample is 16 and the sampling rate is 22.05 kHz.

For the multi-speaker scenario, the Flowtron and WaveGlow are trained on LibriTTS dataset\cite{zen2019libritts}. The LibriTTS corpus consists of 585 hours of speech data at 24 kHz sampling rate from 2,456 speakers and the corresponding text. With the well-trained generative models, given 10,000 random sentences, we generate the corresponding 10,000 audio clips, whose bits per sample is 8 and the sampling rate is 22.05 kHz. To facilitate the steganalysis experiment, we clip the audio clips to 3 seconds.\\
\textbf{Text-to-Image:} The TTI model DALL-E~\cite{ramesh2021zero} is trained on  CUB-200 Birds~\cite{wah2011caltech} that contains around 10k images where each image depicts a single object and there are ten associated captions per image. With the well-trained DALL-E, given 10,000 bird description texts, we generate the corresponding 10,000 spatial images, and the size of the image is $256\times 256\times 3$. For simplicity, the steganography and steganalysis are carried out on the first channel.

\subsubsection{Steganography Algorithm}

To verify the effectiveness of the proposed method, different distortion and different steganographic coding methods are considered.  
DFR~\cite{DBLP:journals/tcsv/ChenZLYZY20} and AACbased~\cite{aacstc} are adopted as the distortion functions for audio steganography, which assign high cost to audio samples which are difficult to predict. HILL~\cite{HILL} and MiPOD~\cite{MiPOD} are adopted as the distortion functions for image steganography, which assign a low cost to those pixels in the texture area and a low cost in the smooth region. STC, SPC and AAC are adopted as steganographic codes. Besides, simulated embedding is also carried out for comparison, which represents the theoretical bound. Since STC frequently fail when the payload is larger than 0.5~\cite{STC}, the payload ranges from 0.1 to 0.5 bps (bit per sample for audio clips) or bpp (bit per pixel for spatial images).  To point out, there are some generative steganography, SteganoGAN~\cite{zhang2019steganogan} and ChatGAN~\cite{tan2021channel}. We have implemented the steganalysis experiments using SRM, and the results show that the detection error rate is near to 0, which is consistent with the results in~\cite{tan2021channel}, meaning that these methods are easy to  detect. Therefore, we do not compare them in the subsequent subsections.

\begin{figure}[t]
	\centering
	\includegraphics[width=3in]{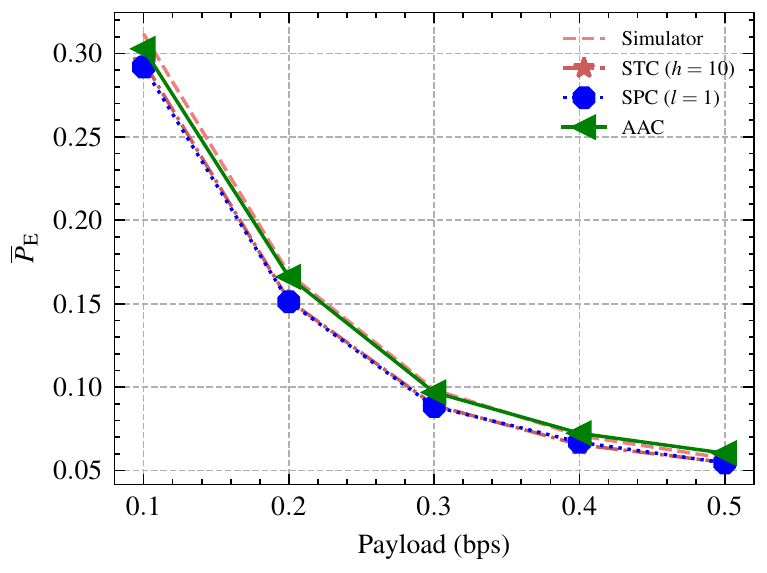}
	\caption{\footnotesize{The average detection error rate $\overline{P}_\textrm{E}$ as a function of payload in bits per sample (bps) for steganographic algorithm payloads ranging from 0.1-0.5 bps against CTM  using \emph{AACbased} distortion on generated \emph{LibriTTS} database.}}
	\label{audio_AAC}
\end{figure}
\begin{figure}[t]
	\centering
	\includegraphics[width=3in]{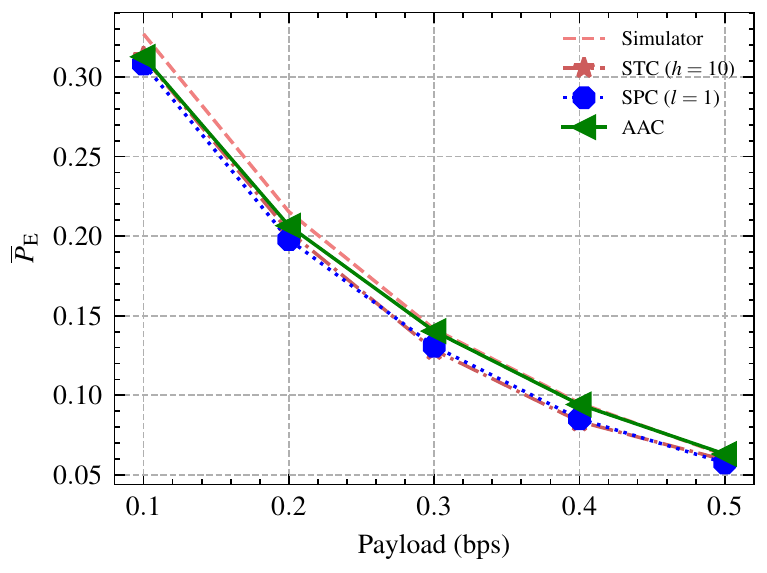}
	\caption{\footnotesize{The average detection error rate $\overline{P}_\textrm{E}$ as a function of payload in bits per sample (bps) for steganographic algorithm payloads ranging from 0.1-0.5 bps against CTM using \emph{DFR} distortion on generated \emph{LibriTTS} database.}}
	\label{audio_DFR}
\end{figure}

\begin{figure}[t]
	\centering
	\includegraphics[width=3in]{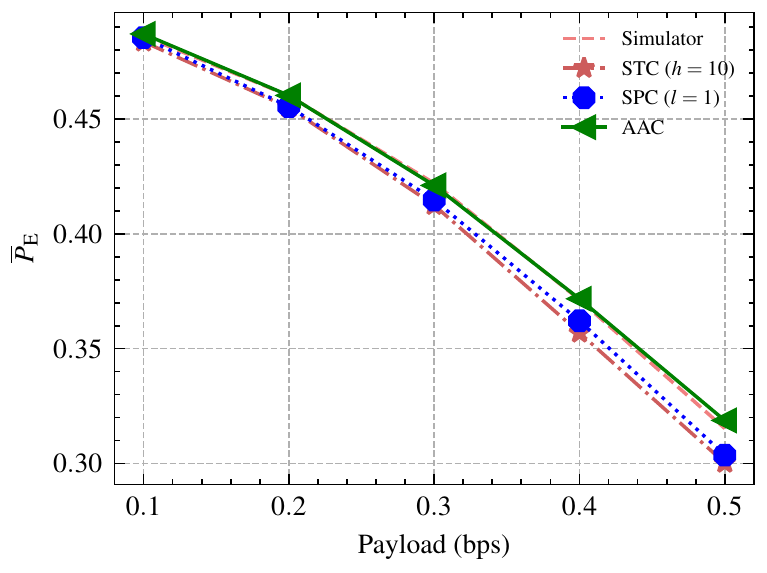}
	\caption{\footnotesize{The average detection error rate $\overline{P}_\textrm{E}$ as a function of payload in bits per sample (bps) for steganographic algorithm payloads ranging from 0.1-0.5 bps against CTM using \emph{AACbased} distortion on generated \emph{LJSpeech} database.}}
	\label{audio_AAC_lj}
\end{figure}
\begin{figure}[t]
	\centering
	\includegraphics[width=3in]{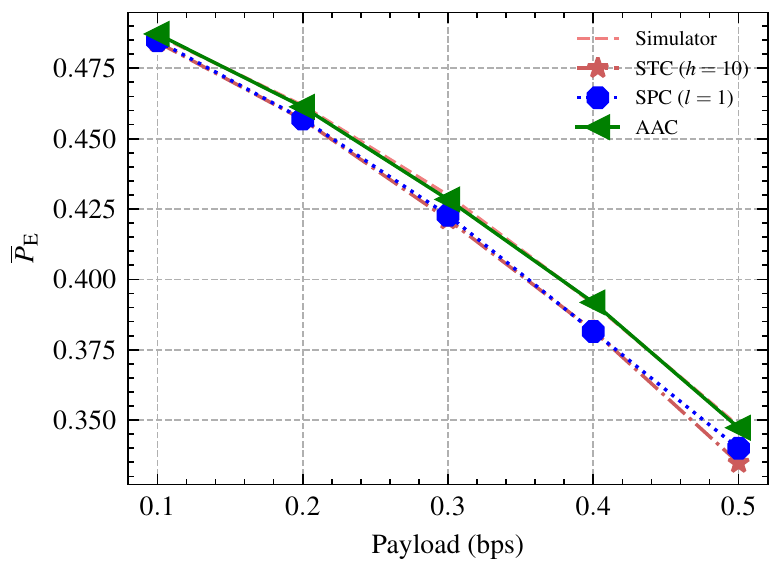}
	\caption{\footnotesize{The average detection error rate $\overline{P}_\textrm{E}$ as a function of payload in bits per sample (bps) for steganographic algorithm payloads ranging from 0.1-0.5 bps against CTM using \emph{DFR} distortion on generated \emph{LJSpeech} database.}}
	\label{audio_DFR_lj}
\end{figure}
\begin{figure}[t]
	\centering
	\includegraphics[width=3in]{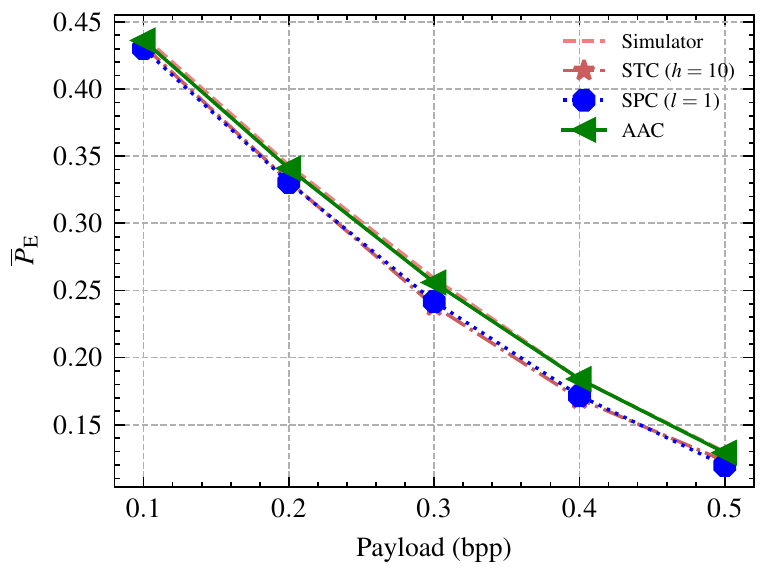}
	\caption{\footnotesize{The average detection error rate $\overline{P}_\textrm{E}$ as a function of payload in bits per pixel (bpp) for steganographic algorithm payloads ranging from 0.1-0.5 bpp against \emph{SRM} using \emph{HILL} distortion on generated \emph{CUB-200 Birds} database.}}
	\label{image_hill_fig}
\end{figure}
\begin{figure}[t]
	\centering
	\includegraphics[width=3in]{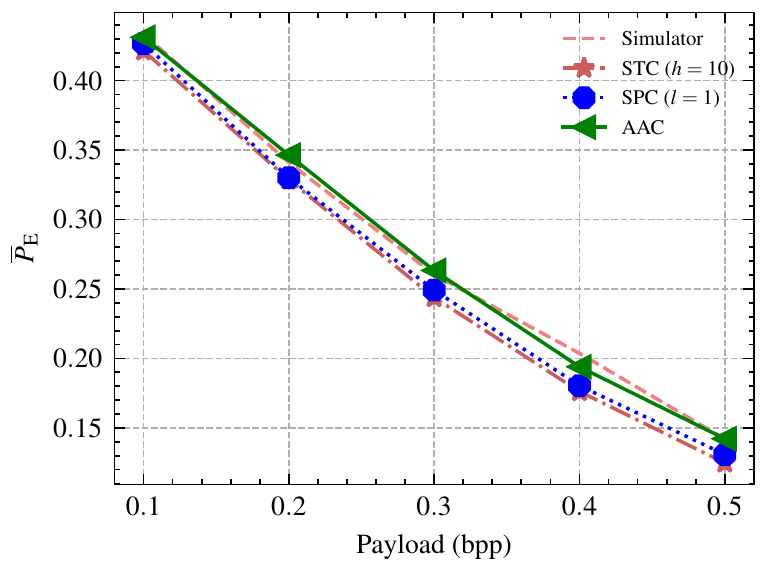}
	\caption{\footnotesize{The average detection error rate $\overline{P}_\textrm{E}$ as a function of payload in bits per pixel (bpp) for steganographic algorithm payloads ranging from 0.1-0.5 bpp against \emph{SRM} using \emph{MiPOD} distortion on generated \emph{CUB-200 Birds} database.}}
	\label{image_mipod_fig}
\end{figure}
\begin{figure}[t]
	\centering
	\includegraphics[width=3in]{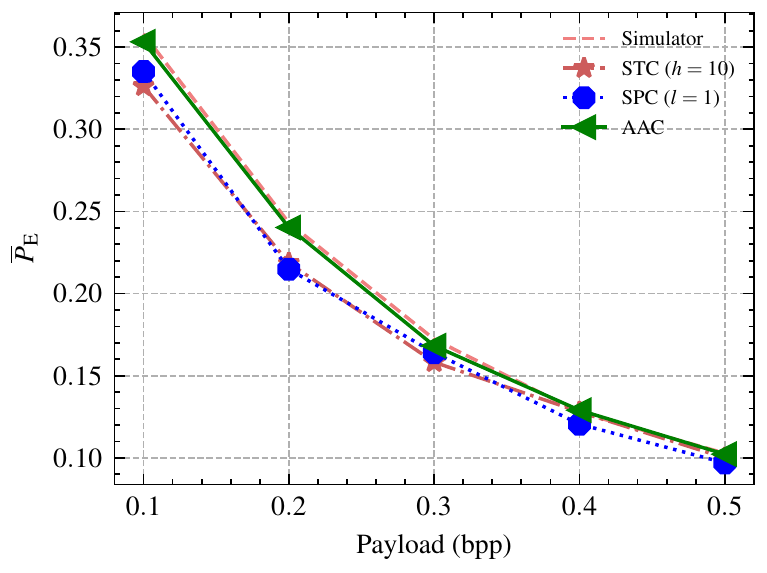}
	\caption{\footnotesize{The average detection error rate $\overline{P}_\textrm{E}$ as a function of payload in bits per pixel (bpp) for steganographic algorithm payloads ranging from 0.1-0.5 bpp against \emph{SRNet} using \emph{HILL} distortion on generated \emph{CUB-200 Birds} database.}}
	\label{image_hill_srnet_fig}
\end{figure}
\begin{figure}[t]
	\centering
	\includegraphics[width=3in]{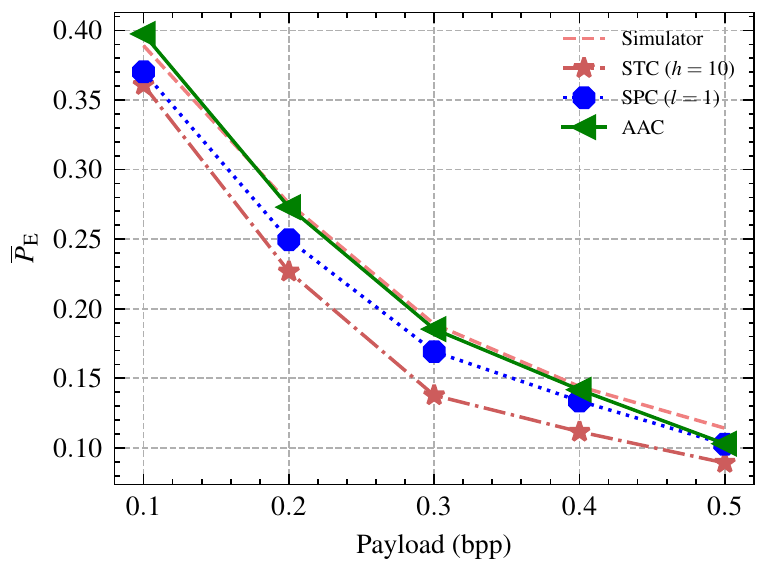}
	\caption{\footnotesize{The average detection error rate $\overline{P}_\textrm{E}$ as a function of payload in bits per pixel (bpp) for steganographic algorithm payloads ranging from 0.1-0.5 bpp against \emph{SRNet} using \emph{MiPOD} distortion on generated \emph{CUB-200 Birds} database.}}
	\label{image_mipod_srnet_fig}
\end{figure}

\subsubsection{Security Evaluation Metric}
In our scenario, it is difficult for an attacker to obtain the same generative model as ours. The attacker also has difficulty training a surrogate model that generates very similar audios or images as our secret generative model. Here, we make a strict assumption that the attacker owns a certain number of image pairs or audio pairs. Then the attacker can use steganalysis for detection.

In regard to audio steganography, the state-of-the-art handcrafted features extracted from both time domain and Mel-spectrogram domain, named CTM, are adopted for detection with ensemble classifier (EC). The ensemble classifier is built on Fisher Linear Discriminator (FLD), which can minimize the total classification error probability under equal priors $P_{\text{E}}=\min_{P_{\text{FA}}} \frac{1}{2}(P_{\text{FA}} +P_{\text{MD}} )$ where $P_{\text{FA}}$ and $P_{\text{MD}}$ are the false-alarm (FA) probability and the missed-detection (MD) probability, respectively. The ultimate security is qualified by average error rate $\bar{P_{\text{E}}}$ averaged over 10 random 50/50 splits of the database, and larger $\bar{P_{\text{E}}}$ means stronger security. Since the deep learning based steganalysis is difficult to detect adaptive audio steganography, as stated in~\cite{zhang2019improving}, we do not implement deep learning based steganalysis for audio. 

As for image steganography, handcrafted feature spatial rich model (SRM~\cite{SRM}) equipped with EC, SRNet~\cite{boroumand2019deep} are adopted. The SRNet is built on TensorFlow.. The optimizer Adamax is used with minibatches of 16 cover-stego pairs. The datasets are divided into training set, validation set, and testing set (7,000, 500, 2500 pairs, respectively). The detection error rate of the testing set is used for evaluating the security of steganographic algorithms.

\subsection{Security Performance and Analysis}

For the text-to-speech task, DFR and AACbased are selected as the distortion function. It can be seen from Figure~\ref{audio_AAC} to  Figure~\ref{audio_DFR_lj} that the AAC-family outperforms STC-family and SPC-family in most cases with respect to different distortions and different datasets. Additionally, the detection error rate is also approaching the theoretical bound. The biggest improvement is near to 2\%.

For the text-to-image task, HILL and MiPOD are selected as the distortion function. Figure~\ref{image_hill_fig} and Figure~\ref{image_mipod_fig} show the security performance of the different coding methods against SRM. In most cases, AAC-family outperforms STC-family as well as SPC-family and is closer to the theoretical bound. The results verify the effectiveness of the CRS framework applied to text-to-image tasks. Numerically, AAC-family outperforms STC-family by at most 2\% for HILL. As for MiPOD, the tendency of the improvement is similar. Figure~\ref{image_hill_srnet_fig} and Figure~\ref{image_mipod_srnet_fig} show the security performance against SRNet. The tendency of the results is similar to that against SRM. In most cases, AAC-family outperforms STC-family and SPC-family. 


To confirm the statistical significance of the improvement, we apply a $t$-test to evaluate the statistical significance of the results. The hypotheses are
\begin{equation}
  H_{0}:\mu1 \leq \mu2 ; H_{1}:\mu1 > \mu2
\end{equation}
where $\mu1$ and $\mu2$ are the mean values of the average detection error of the improved method (AAC) and the original method (STC).
The statistic $t$ is calculated as follows:
\begin{equation}
  t=\frac{\mu_{1}-\mu_{2}}{ \sqrt{S_{w}(\frac{1}{n_{1}}+\frac{1}{n_{2}})}} 
 \end{equation}
 where 
 \begin{equation}
  S_{w}=\frac{1}{n_{1}+n_{2}-2}\left[\left(n_{1}-1\right) S_{1}^{2}+\left(n_{2}-1\right) S_{2}^{2}\right] 
 \end{equation}
$n_1$ and $n_2$ are the numbers of testing times, and $S_1$ and $S_2$ are the standard deviations of the original and improved algorithms, respectively. By looking up the $t$-score table of the standard normal distribution, the corresponding $t$-value can be obtained. A lower $t$-value indicates a higher probability that $H_{0}$ holds. If the $t$-value is larger than a threshold, $H_{0}$ is rejected, and the improvement is deemed statistically signiﬁcant and reliable.

The signiﬁcance level for the test is set to $t_{0.05}(n_1+n_2-2)$. As shown in Table~\ref{tab1} to Table~\ref{tab666}, under different payloads and steganographic schemes, in most cases, the test statistic $t$-values are larger than the corresponding quantile $t_{0.05}(18)=1.734$, which implies that the detection improvement with respect to STC has statistical signiﬁcance. 

\subsection{Randomness Analysis}
The parameters of the neural network in our CRS framework are seen as secret keys, which are only known by the sender and the receiver. The attacker cannot obtain the identical model. 
Note that, in the process of training a neural network, there are multiple stages where randomness is used, such that the previous settings can be established. The randomness operations include:
\begin{itemize}
	\item Random initialization of weights of the network before the training starts, e.g., RandomNormal, TruncatedNormal and RandomUniform.
	\item Regularization, e.g., dropout, which involves randomly dropping nodes in the network while training.
	\item Optimization processes like stochastic gradient descent, RMSProp or Adam also include random initializations.
\end{itemize}
These operations can be controlled by a seed. Therefore, the seed can be regarded as the secret key of the method under the CRS framework. Meanwhile, the choice of the seed has little impact on the performance of the neural network, which indicates that the behavior does not arouse suspicion from the attacker.

\subsection{Performance of Distortion-Sort Steganography}
To verify the inference we concluded in Section \ref{sec3}, steganalysis on distortion-sort steganography is carried out. Two representative strategies are  considered:
\begin{itemize}
    \item ``SORT'': select $L$ (message length) cover elements as the final cover, set their corresponding distortion as the final distortion,
    and use simulate embedding for message embedding.
    \item ``SORT-EVEN'': select $L/\log_23$  cover elements as the final cover, set the modification distortion to zero ($p_{-1}=p_{+1}=p_{0}=\frac{1}{3}$),
    and use simulate embedding for message embedding.
\end{itemize}
Table~\ref{dis-sort} presents the results, where HILL is the distortion function and SRM is the steganalyzer. HILL performs far better than HILL-SORT and HILL-SORT-EVEN, indicating the the distortion-sort steganography cannot surpass minimize distortion steganography, which validates our conclusion in Section~\ref{sec3}.

\begin{table}
\centering
\caption{The detection errors $\overline{P}_{\text{E}}$ of distortion-sort steganography with different payloads (0.1 - 0.5 bpp) against SRM.}
\label{dis-sort}
\begin{tabular}{l|ccccc} 
\hline
Method              & 0.1             & 0.2             & 0.3             & 0.4             & 0.5              \\ 
\hline
HILL           & \textbf{0.4391} & \textbf{0.3433} & \textbf{0.2590} & \textbf{0.1839} & \textbf{0.1298}  \\
HILL-SORT      & 0.3750          & 0.2760          & 0.1976          & 0.1428          & 0.1004           \\
HILL-SORT-EVEN & 0.3910          & 0.2725          & 0.1844          & 0.1277          & 0.0861                \\
\hline
\end{tabular}
\end{table}




\subsection{Time Complexity Comparison}
The time complexity of the embedding process of STC, SPC, and AAC are $O(2^hn)$~\cite{STC}, $O(l\cdot n \log_2 n)$~\cite{tal2015list} and $O(n)$~\cite{ramabadran1990coding}. Therefore, arithmetic coding has certain advantages at the sender-end. Since the implementation details are different, such as the source codes of STC have adopted Intel Instruction Set for faster implication, we do not compare the exact running time here.
Regarding message extraction, there are additional operations in our CRS framework, such as audio generation and modification probability calculation, while STC and SPC only need to calculate the syndrome function. Therefore, CRS framework is more time-consuming for receivers. Intuitively, we present the average running time of different parts of the proposed CRS framework in Table~\ref{runningtime}. It can be seen that the time cost of the whole process is small and the
processes of embedding and extraction using arithmetic coding are very efficient with respect to generating cover.

\begin{table}[h]
	\centering
	\caption{The running time of different parts of the proposed CRS framework. (GPU: Waveglow: Nvidia  3090, DALL-E: Nvidia A100, \\
		CPU: Intel (R) Xeon (R) CPU E5-2670 @2.30 GHz) }
	\begin{tabular}{@{}cccc@{}}
		\toprule
		      Generative Models & Generation (s) & Embedding (s)  & Extraction (s) \\ \midrule
		 WaveGlow~\cite{prenger2019waveglow}          & 1.883              &  0.0153                   & 0.0163                    \\
		 DALL-E~\cite{ramesh2021zero}            & 0.1593             & 0.0156                    & 0.0149                   \\ \bottomrule
		\label{runningtime}
	\end{tabular}
\end{table}

\begin{table*}
	\centering
	\caption{The average detection errors $\overline{P}_{\text{E}}$ of different coding methods with \emph{HILL} and  \emph{MiPOD} distortion on the generated \emph{CUB-200 Birds} dataset.}
	\label{tab1}
	\begin{tabular}{|c|c|c|c|c|c|c|} 
		\hline
		\multicolumn{2}{|c|}{\diagbox{Method}{Payload}} & 0.1                                  & 0.2                                  & 0.3                                  & 0.4                                  & 0.5                                   \\ 
		\hline
		\multirow{5}{*}{HILL}  & SIMU                   & 0.4391 (+/- 0.0025)                  & 0.3433 (+/- 0.0066)                  & 0.2590 (+/- 0.0061)                  & 0.1839 (+/- 0.0040)                  & 0.1298 (+/- 0.0020)                   \\
		& STC                    & 0.4323 (+/- 0.0033)                  & 0.3312 (+/- 0.0055)                  & 0.2384 (+/- 0.0046)                  & 0.1691 (+/- 0.0029)                  & 0.1229 (+/- 0.0036)                   \\
		& SPC                    & 0.4301 (+/- 0.0043)                  & 0.3304 (+/- 0.0037)                  & 0.2417 (+/- 0.0033)                  & 0.1716 (+/- 0.0042)                  & 0.1196 (+/- 0.0038)                   \\
		& AAC                    & \uline{\textbf{0.4362 (+/- 0.0054)}}                  & \uline{\textbf{0.3409 (+/- 0.0047)}} & \uline{\textbf{0.2559 (+/- 0.0041)}} & \uline{\textbf{0.1840 (+/- 0.0054)}} & \uline{\textbf{0.1291 (+/- 0.0045)}}  \\
		& $t$-value                & 1.9488           & 4.2398           &8.9809         & 7.6872         & 3.4022                                \\ 
		\hline
		\multirow{5}{*}{MiPOD} & SIMU                   & 0.4336 (+/- 0.0051)                  & 0.3411 (+/- 0.0074)                  & 0.2604 (+/- 0.0050)                  & 0.2037 (+/- 0.0035)                  & 0.1421 (+/- 0.0030)                   \\
		& STC                    & 0.4215 (+/- 0.0034)                  & 0.3286 (+/- 0.0048)                  & 0.2439 (+/- 0.0034)                  & 0.1762 (+/- 0.0031)                  & 0.1248 (+/- 0.0033)                   \\
		& SPC                    & 0.4267 (+/- 0.0048)                  & 0.3301 (+/- 0.0057)                  & 0.2492 (+/- 0.0039)                  & 0.1804 (+/- 0.0035)                  & 0.1309 (+/- 0.0030)                   \\
		& AAC                    & \uline{\textbf{0.4312 (+/- 0.0039)}} & \uline{\textbf{0.3464 (+/- 0.0061)}} & \uline{\textbf{0.2633 (+/- 0.0060)}} & \uline{\textbf{0.1940 (+/- 0.0038)}} & \uline{\textbf{0.1422 (+/- 0.0033)}}  \\
		& $t$-value                & 5.9285                               & 7.2517                               & 8.8957                               & 11.477                               & 11.790                                \\
		\hline
	\end{tabular}
\end{table*}

\begin{table*}
	\centering
	\caption{The average detection errors $\overline{P}_{\text{E}}$ of different coding methods with \emph{AACbased} and \emph{DFR} distortion on generated \emph{LibriTTS} dataset.}
	\label{HILL_PE2}
	\begin{tabular}{|c|c|c|c|c|c|c|} 
		\hline
		\multicolumn{2}{|c|}{\diagbox{Method}{Payload}} & 0.1                                  & 0.2                                  & 0.3                                  & 0.4                                  & 0.5                                   \\ 
		\hline
		\multirow{5}{*}{AACbased}  &   SIMU                          & 0.3117 (+/- 0.0022) & 0.1680 (+/- 0.0030) & 0.0985 (+/- 0.0023) & 0.0709 (+/- 0.0032) & 0.0574 (+/- 0.0022)  \\ 
		&STC                           & 0.2950 (+/- 0.0044) & 0.1520 (+/- 0.0023) & 0.0892 (+/- 0.0019) & 0.0655 (+/- 0.0020) & 0.0549 (+/- 0.0020)  \\ 
		&SPC                           & 0.2919 (+/- 0.0038) & 0.1512 (+/- 0.0020) & 0.0884 (+/- 0.0014) & 0.0670 (+/- 0.0022) & 0.0546 (+/- 0.0024)  \\ 	
		&AAC                           & \underline{\bf{0.3028 (+/- 0.0028)}} & \underline{\bf{0.1661 (+/- 0.0021)}} & \underline{\bf{0.0969 (+/- 0.0023)}} & \underline{\bf{0.0724 (+/- 0.0023)}} & \underline{\bf{0.0604 (+/- 0.0026)}}  \\
		
		& $t$-value                & 4.7294        & 14.3164        & 8.1620         & 7.1588          & 5.3022                              \\ 
		\hline
		\multirow{5}{*}{DFR} &   SIMU                          & 0.3271 (+/- 0.0020) & 0.2154 (+/- 0.0029) & 0.1416 (+/- 0.0026) & 0.0951 (+/- 0.0028) & 0.0624 (+/- 0.0021)  \\ 
		&STC                           & 0.3125 (+/- 0.0036) & 0.2029 (+/- 0.0025) & 0.1284 (+/- 0.0021) & 0.0835 (+/- 0.0021) & 0.0593 (+/- 0.0019)  \\ 
		&SPC                           & 0.3082 (+/- 0.0035) & 0.1979 (+/- 0.0023) & 0.1308 (+/- 0.0017) & 0.0852 (+/- 0.0024) & 0.0575 (+/- 0.0023)  \\ 	
		&AAC                           & 0.3127 (+/- 0.0025)& \underline{\bf{0.2066 (+/- 0.0023)}} & \underline{\bf{0.1403 (+/- 0.0026)}} & \underline{\bf{0.0942 (+/- 0.0029)}} & \underline{\bf{0.0631 (+/- 0.0022)}}  \\
				& $t$-value                & 0.1443                            & 3.4443                              & 11.2595                              & 9.4502                             & 4.1339                               \\
		\hline
	\end{tabular}
\end{table*}

\begin{table*}
	\centering
	\caption{The average detection errors $\overline{P}_{\text{E}}$ of different coding methods with \emph{AACbased} and \emph{DFR} distortion on generated \emph{LJSpeech} dataset.}
	\label{tab666}
	\begin{tabular}{|c|c|c|c|c|c|c|} 
		\hline
		\multicolumn{2}{|c|}{\diagbox{Method}{Payload}} & 0.1                                  & 0.2                                  & 0.3                                  & 0.4                                  & 0.5                                   \\ 
		\hline
		\multirow{5}{*}{AACbased}  &     SIMU                          & 0.4862 (+/- 0.0015) & 0.4602 (+/- 0.0022) & 0.4221 (+/- 0.0027) & 0.3713 (+/- 0.0018) & 0.3152 (+/- 0.0025)  \\ 
		&STC                           & 0.4837 (+/- 0.0019) & 0.4550 (+/- 0.0019) & 0.4125 (+/- 0.0033) & 0.3569 (+/- 0.0017) & 0.3002 (+/- 0.0034)  \\ 
		&SPC                           & 0.4855 (+/- 0.0029) & 0.4555 (+/- 0.0023) & 0.4148 (+/- 0.0025) & 0.3621 (+/- 0.0030) & 0.3036 (+/- 0.0034)  \\ 	
		&AAC                           & \underline{\bf{0.4871 (+/- 0.0016)}} & \underline{\bf{0.4603 (+/- 0.0016)}} & \underline{\bf{0.4211 (+/- 0.0034)}} & \underline{\bf{0.3717 (+/- 0.0023)}} & \underline{\bf{0.3187 (+/- 0.0024)}}  \\
		
		& $t$-value                & \multicolumn{1}{c}{4.3285}           & \multicolumn{1}{c}{6.7474}           & \multicolumn{1}{c}{5.7397}           & \multicolumn{1}{c}{16.3638}           & 14.0571                                \\ 
		\hline
		\multirow{5}{*}{DFR} &     SIMU                          & 0.4868 (+/- 0.0014) & 0.4618 (+/- 0.0014) & 0.4299 (+/- 0.0025) & 0.3914 (+/- 0.0030) & 0.3479 (+/- 0.0030)  \\ 
		&STC                           & 0.4843 (+/- 0.0010) & 0.4568 (+/- 0.0015) & 0.4213 (+/- 0.0014) & 0.3818 (+/- 0.0025) & 0.3348 (+/- 0.0022)  \\ 
		&SPC                           & 0.4848 (+/- 0.0009) & 0.4570 (+/- 0.0021) & 0.4228 (+/- 0.0016) & 0.3815 (+/- 0.0033) & 0.3400 (+/- 0.0026)  \\ 	
		&AAC                           & \underline{\bf{0.4872 (+/- 0.0013)}} & \underline{\bf{0.4613 (+/- 0.0021)}} & \underline{\bf{0.4284 (+/- 0.0018)}} & \underline{\bf{0.3918 (+/- 0.0015)}} & \underline{\bf{0.3473 (+/- 0.0019)}}  \\
				& $t$-value                & 6.5555                           & 6.4944                             & 11.9261                             & 16.0529                             & 20.1253                         \\
		\hline
	\end{tabular}
\end{table*}

\section{Conclusion}
\label{sec6}
In this paper, we have proposed an effective cover reproducible steganography framework based on generative models. The generative models and the robust semantic of the generated media make it possible for the receiver to reconstruct cover. Under this circumstance, we introduce arithmetic coding for message embedding and extraction, and the superior performance has also been proven. Besides, two instances are designed under our proposed framework based on text-to-speech and text-to-image tasks. Experimental results show that the steganographic methods under cover reproducible steganography framework are more secure than traditional STC and SPC methods. Furthermore, we also point out that distortion-sort stenography is not a considerable solution. 

The generative model not only introduces a new data environment, but also provides a new steganography framework. The proposed method can only work on lossless channels. In the future, we will delve into how to improve robust steganography based on generative models.

\ifCLASSOPTIONcompsoc
  \section*{Acknowledgments}
\else
  \section*{Acknowledgment}
\fi

The authors would like to thank DDE Laboratory of SUNY Binghamton for sharing the source code of steganography on the webpage (http://dde.binghamton.edu/download/).

\ifCLASSOPTIONcaptionsoff
  \newpage
\fi

\bibliographystyle{IEEEtran}
\bibliography{wifs}

\begin{IEEEbiography}[{\includegraphics[width=1in,height=1.25in,clip,keepaspectratio]{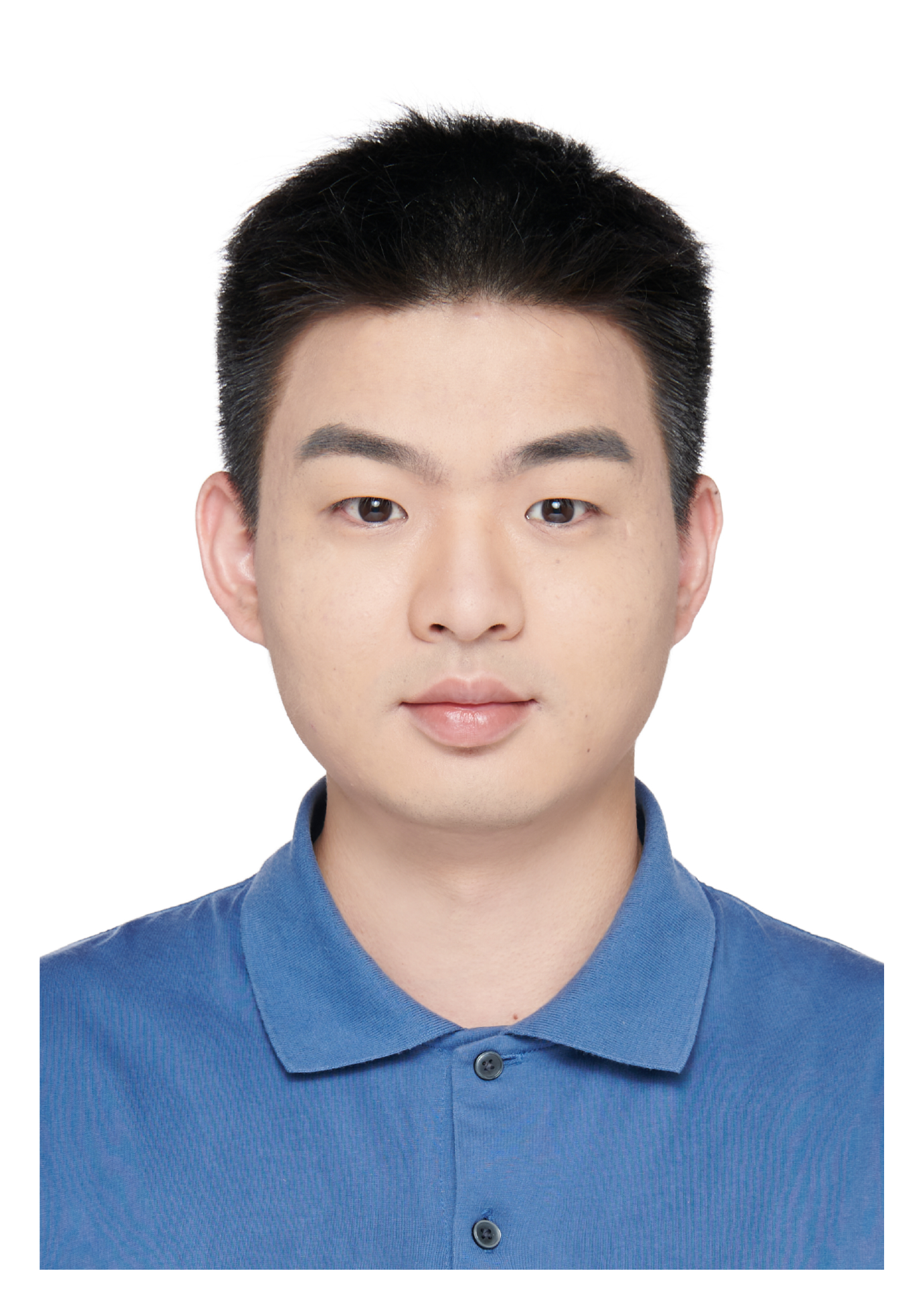}}]{Kejiang Chen}
    received his B.S. degree in 2015 from Shanghai University (SHU) and a Ph.D. degree in 2020 from the University of Science and Technology of China (USTC). Currently, he is a Researcher Associate at the University of Science and Technology of China. His research interests include information hiding, image processing and deep learning.
    \end{IEEEbiography}

    \begin{IEEEbiography}[{\includegraphics[width=1in,height=1.25in,clip,keepaspectratio]{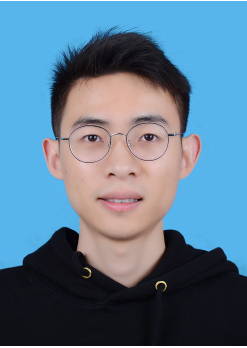}}]{Hang Zhou}
    received his B.S. degree in 2015 from Shanghai University (SHU) and a Ph.D. degree in 2020 from the University of Science and Technology of China (USTC). Now he is a postdoctoral researcher at Simon Fraser University.
    His research interests include computer graphics, multimedia security and deep learning.
    \end{IEEEbiography} 
	
    \begin{IEEEbiography}[{\includegraphics[width=1in,height=1.25in,clip,keepaspectratio]{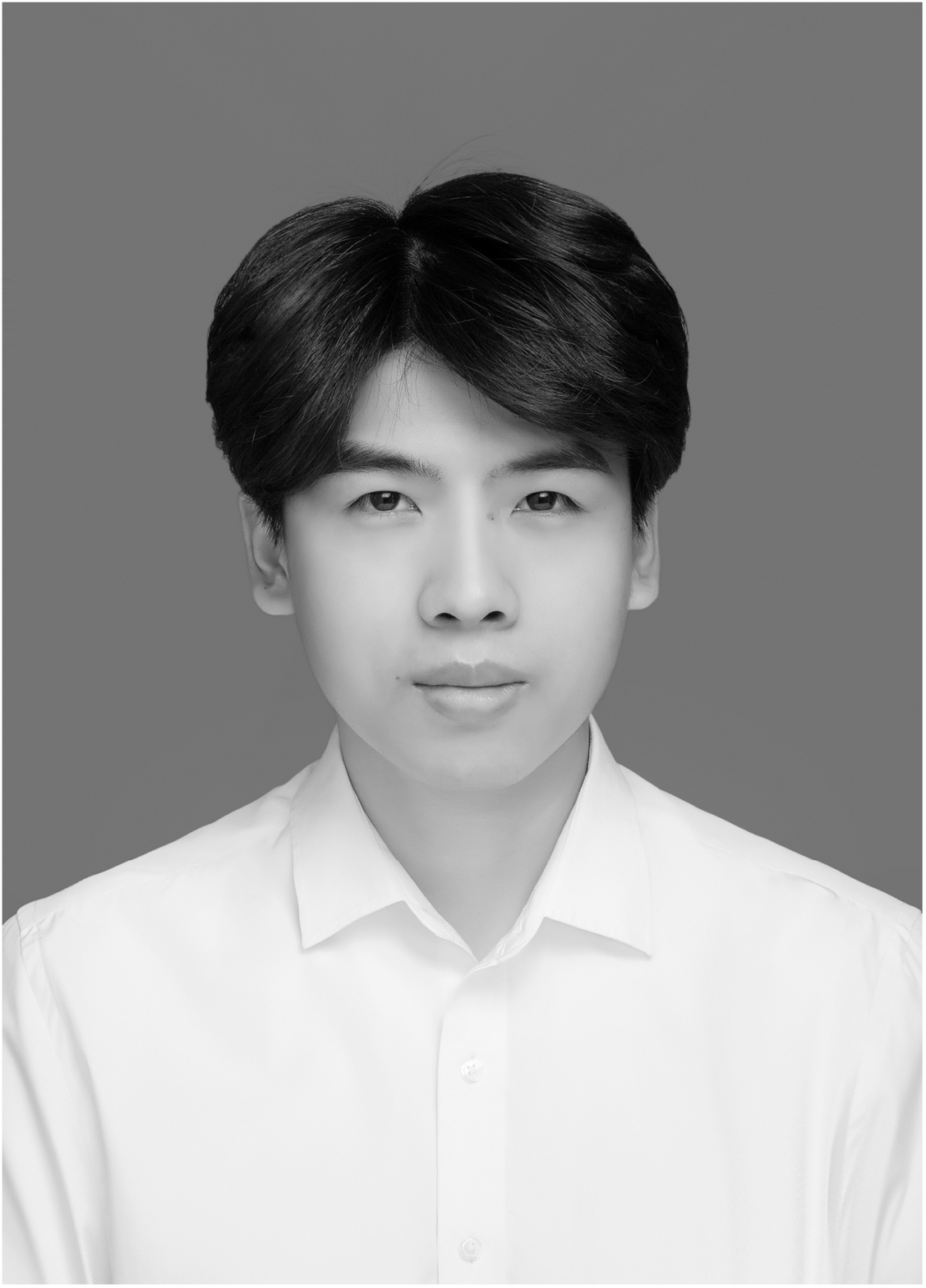}}]{Yaofei Wang}
    received the  the B.S. degree from the Southwest Jiaotong University in 2017 and the Ph.D. degree from the University of Science and Technology of China in 2022. Currently, he is an Associate Professor at the Hefei University of Technology. His research interests include information hiding, image processing, and deep learning.
    \end{IEEEbiography} 

	\begin{IEEEbiography}[{\includegraphics[width=1in,height=1.25in,clip,keepaspectratio]{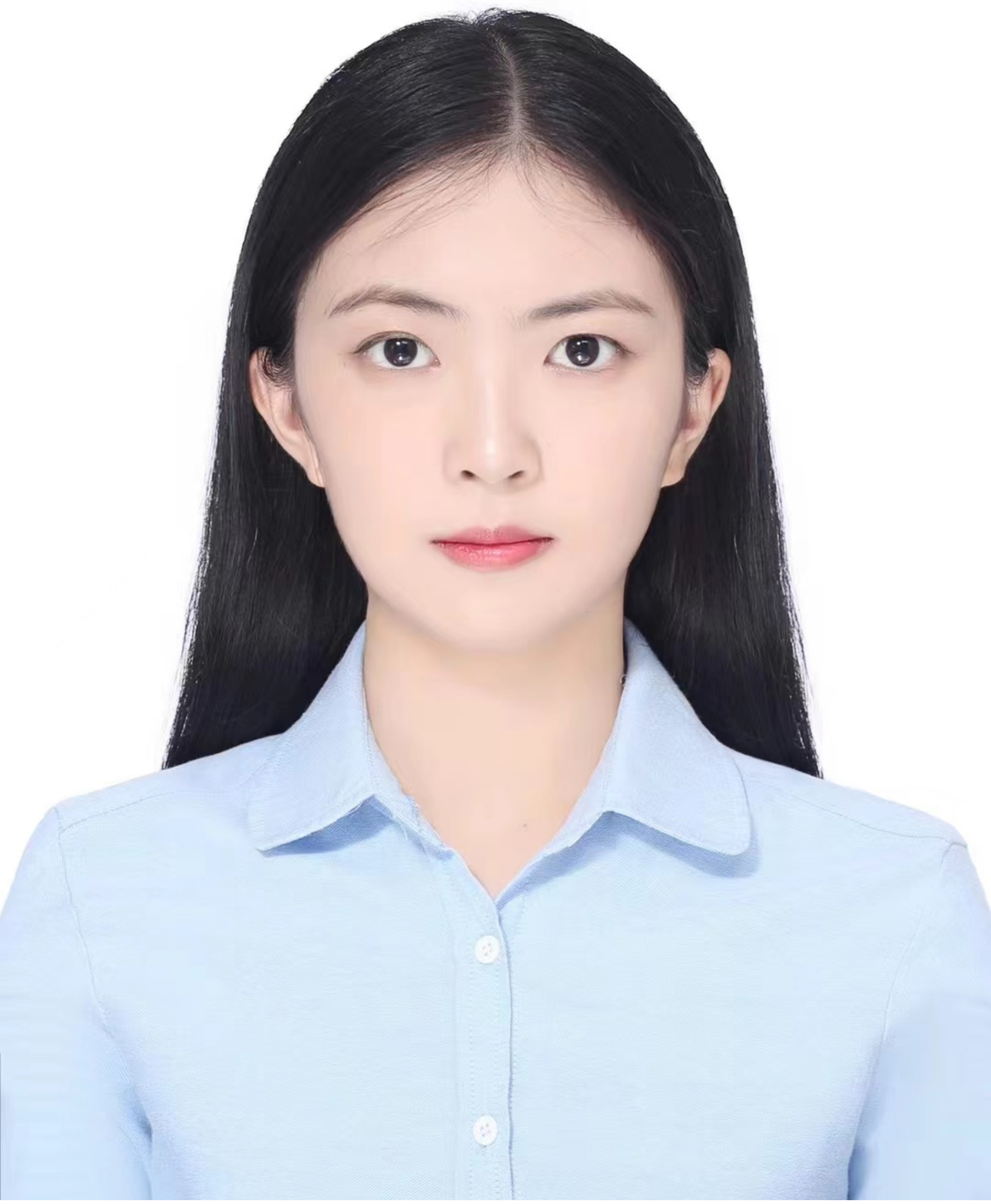}}]{Menghan Li}
    received the B.S. degree and M.S. degree in 2018 and 2021, respectively, from the University of Science and Technology of China. Her research interests include information hiding.
	\end{IEEEbiography}

    
    \begin{IEEEbiography}[{\includegraphics[width=1in,height=1.25in,clip,keepaspectratio]{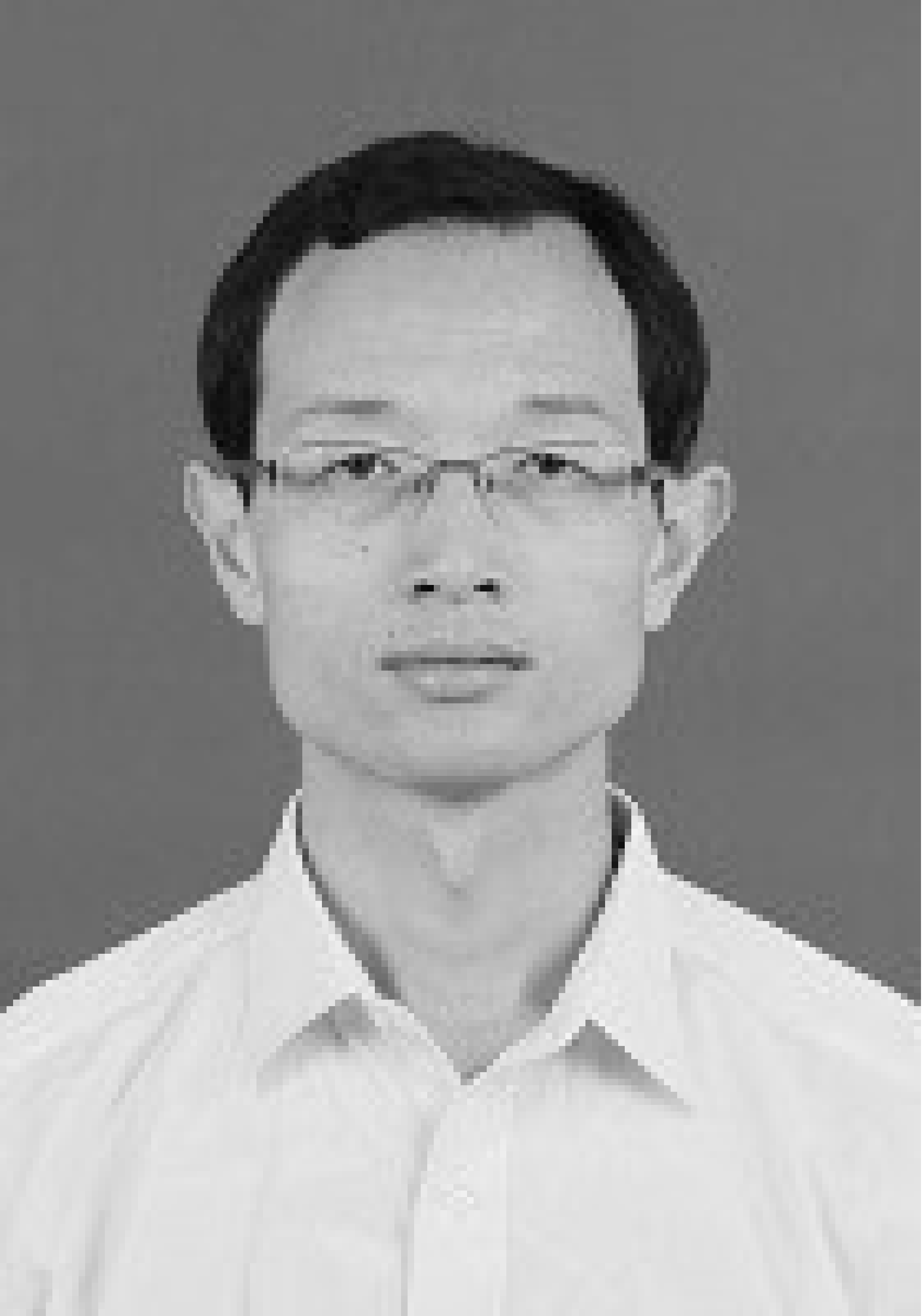}}]{Weiming Zhang}
    received his M.S. degree and Ph.D. degree in 2002 and 2005, respectively, from the Zhengzhou Information Science and Technology Institute, P.R. China. Currently, he is a professor with the School of Information Science and Technology, University of Science and Technology of China. His research interests include information hiding and multimedia security.
    \end{IEEEbiography}
    
    \begin{IEEEbiography}[{\includegraphics[width=1in,height=1.25in,clip,keepaspectratio]{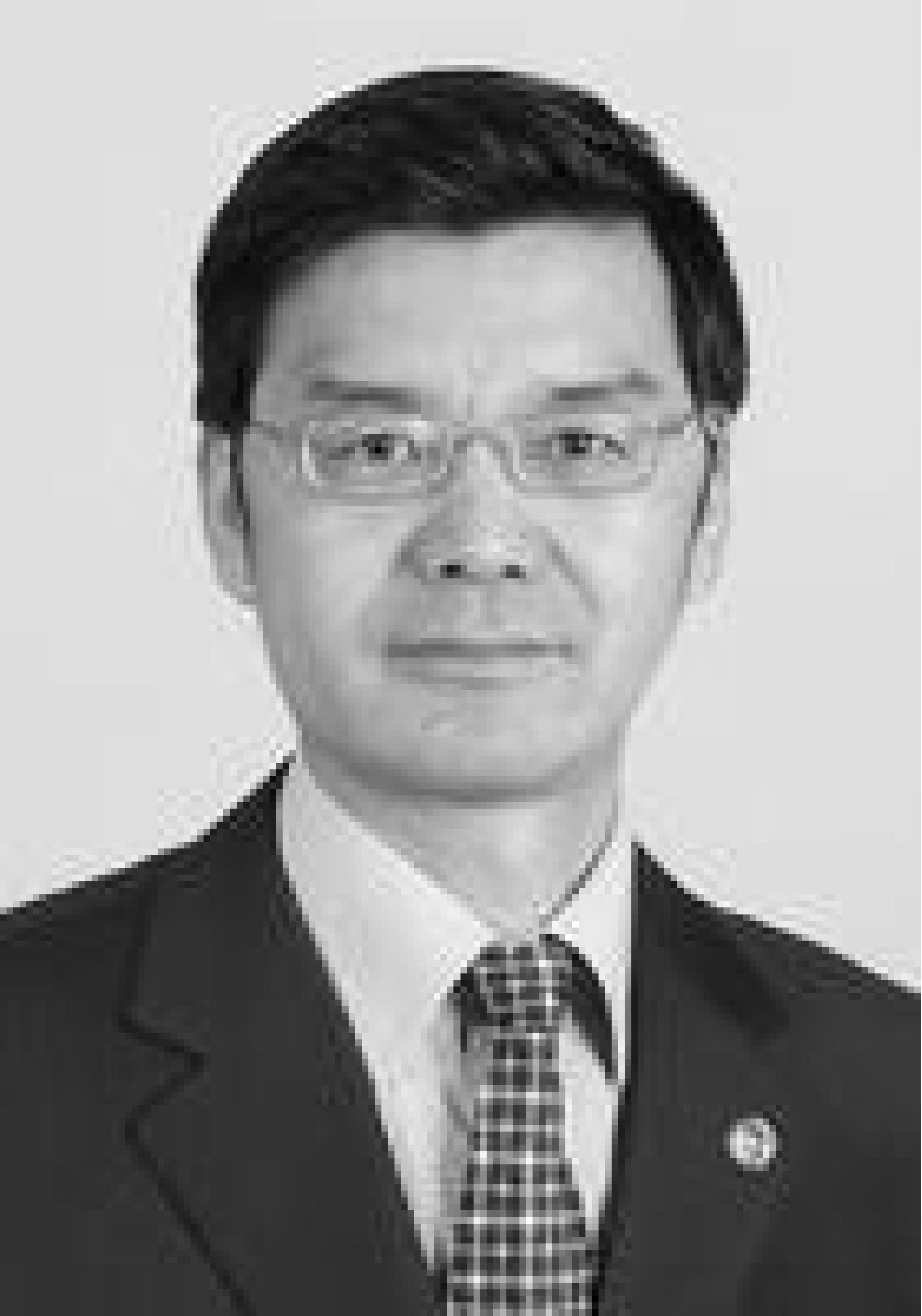}}]{Nenghai Yu}
    received his B.S. degree in 1987 from Nanjing University of Posts and Telecommunications, an M.E. degree in 1992 from Tsinghua University and a Ph.D. degree in 2004 from the University of Science and Technology of China, where he is currently a professor. His research interests include multimedia security, multimedia information retrieval, video processing and information hiding.
    \end{IEEEbiography}



%








\end{document}